\documentclass[aps, twocolumn, superscriptaddress,]{revtex4} 

\usepackage{amsmath}
\usepackage{empheq}
\usepackage[parfill]{parskip}
\usepackage{graphicx}
\usepackage[active]{srcltx}
\usepackage{color}
\usepackage{array}
\newcolumntype{P}[1]{>{\centering\arraybackslash}p{#1}}
\usepackage{amsfonts}
\usepackage{dsfont}
\usepackage{float}
\usepackage{subcaption}
\begin{document}
\setlength{\parindent}{0.5cm}

\title{Quantifying the sensing power of crowd-sourced vehicle fleets}

\author{Kevin P. O'Keeffe}
\affiliation{Senseable City Lab, Massachusetts Institute of Technology, Cambridge, MA 02139} 

\author{Amin Anjomshoaa}

\affiliation{Senseable City Lab, Massachusetts Institute of Technology, Cambridge, MA 02139} 

\author{Steven H. Strogatz}
\affiliation{Department of Mathematics, Cornell University, Ithaca, NY 14853} 

\author{Paolo Santi}
\affiliation{Senseable City Lab, Massachusetts Institute of Technology, Cambridge, MA 02139}
\affiliation{Istituto di Informatica e Telematica del CNR, Pisa, ITALY}

\author{Carlo Ratti}
\affiliation{Senseable City Lab, Massachusetts Institute of Technology, Cambridge, MA 02139}

\maketitle

\textbf{Sensors can measure air quality, traffic congestion, and other aspects of urban environments. The fine-grained diagnostic information they provide could help urban managers to monitor a city's health \cite{lane2008urban,cuff2008urban,rashed2010remote, dutta2009common}. Recently, a `drive-by' paradigm has been proposed in which sensors are deployed on third-party vehicles, enabling wide coverage at low cost \cite{driveby1,driveby2,driveby3,driveby4}. Research on drive-by sensing has mostly focused on sensor engineering \cite{driveby_network1, driveby_network2, driveby_data1, driveby_data2, driveby_data3}, but a key question remains unexplored:  How many vehicles would be required to adequately scan a city? Here, we address this question by analyzing the sensing power of a taxi fleet. Taxis, being numerous in cities and typically equipped with some sensing technology (e.g. GPS), are natural hosts for the sensors. Our strategy is to view drive-by sensing as a spreading process, in which the area of sensed terrain expands as sensor-equipped taxis diffuse through a city's streets. In tandem with a simple model for the movements of the taxis, this analogy lets us analytically determine the fraction of a city's street network sensed by a fleet of taxis during a day. Our results agree with taxi data obtained from nine major cities, and reveal that a remarkably small number of taxis can scan a large number of streets. This finding appears to be universal, indicating its applicability to cities beyond those analyzed here. Moreover, because taxi motions combine randomness and regularity (passengers' destinations being random, but the routes to them being deterministic), the spreading properties of taxi fleets are unusual; in stark contrast to random walks, the stationary densities of our taxi model obey Zipf's law, consistent with the empirical taxi data. Our results have direct utility for town councilors, smart-city designers, and other  urban decision makers.} 

\begin{figure}
  \includegraphics[width=\columnwidth]{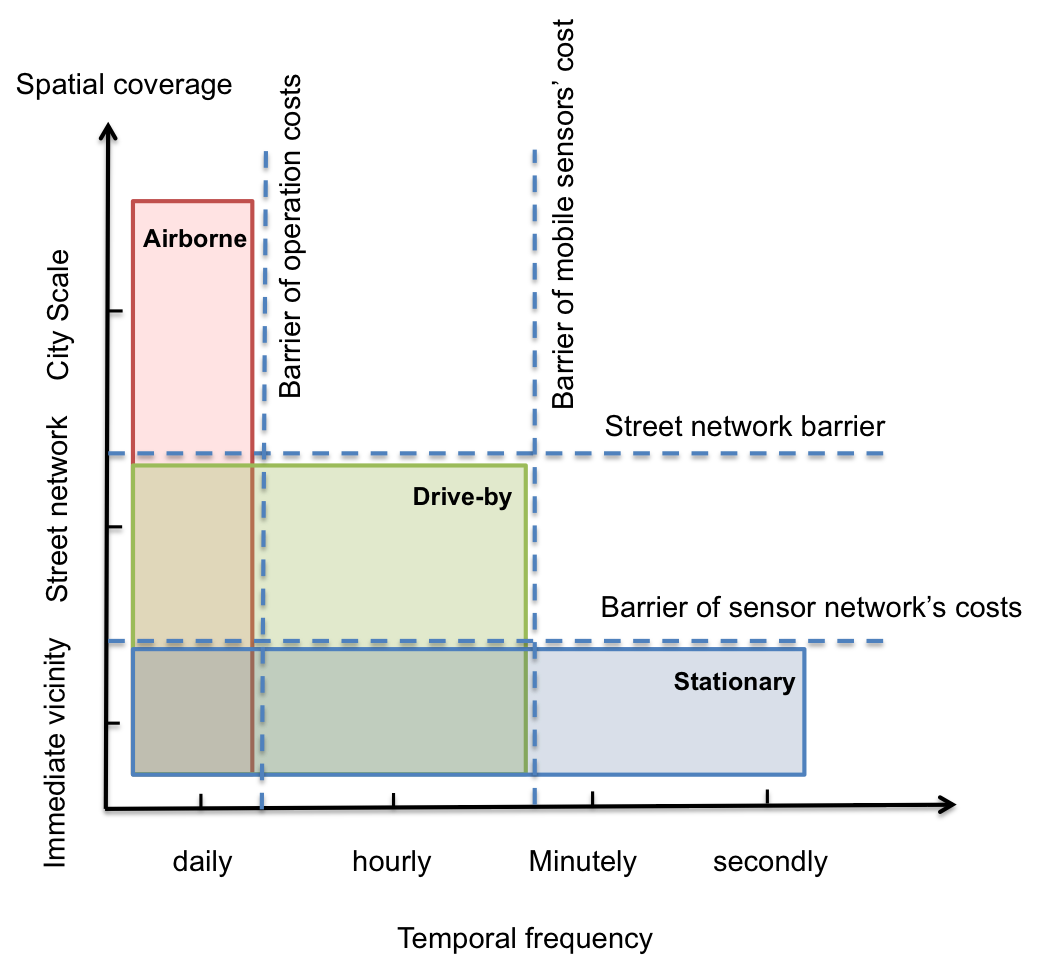}
  \caption{\textbf{Comparison of different sensing methods}. Airborne sensors, such as satellites, provide good spatial coverage, but their temporal coverage is limited to the time interval when the sensors pass over the location being sensed. Conversely, stationary sensors collect data for long periods of time, but have limited spatial range. Drive-by sensing offers some advantages of both methods. By utilizing host vehicles as `data mules,' drive-by sensing offers a cheap, scalable, and sustainable way to accurately monitor cities in both space and time.}
  \label{sensing_types}
\end{figure}

Traditional approaches to urban sensing fall into two main categories (Fig.~\ref{sensing_types}), each of which has limitations \cite{lane2008urban,cuff2008urban,rashed2010remote}. At one extreme, airborne sensors such as satellites scan wide areas, but only during certain time windows. At the other extreme, stationary sensors collect data over long periods of time, but with limited spatial range. Drive-by sensing addresses the weakness in both these methods and offers good coverage in both space and time. In particular, mounting sensors on crowd-sourced urban vehicles, such as cars, taxis, buses, or trucks, enables them to scan the wide areas traversed by their hosts, allowing air pollution, road quality, and other urban metrics to be monitored at fine-scale spatiotemporal resolutions.

\begin{figure*}
  \includegraphics[width= 2\columnwidth]{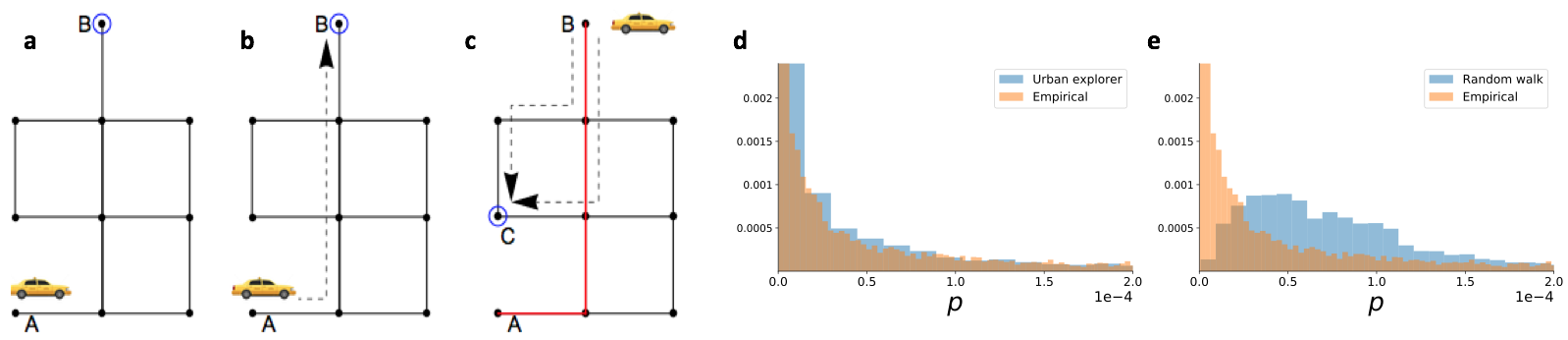}
  \caption{\textbf{Urban explorer process}. Panels (a)-(c) show a schematic of the urban explorer process. (a) A taxi picks up a passenger at node $A$. Then a destination node $B$ (blue circle) is randomly chosen. (b) The shortest path between $A$ and $B$ is taken (dashed arrow). No edges have yet been sensed. (c) After the edges connecting $A$ and $B$ have been traversed by the sensor-equipped taxi, they become `sensed,' which we denote by coloring them red. Now at $B$, the taxi proceeds to its next pickup at, say, $C$. There are two shortest paths connecting $B$ and $C$, so one is chosen at random. This process then repeats. (d) Distribution of street segment popularities $p$ predicted by the urban explorer process (blue histogram) agree with empirical data from Manhattan (brown histogram). (e) By contrast, a random walk model of taxi movement incorrectly predicts a skewed, unimodal distribution of street segment popularities, in qualitative disagreement with the data. For panels (d) and (e) the (directed) Manhattan street network on which the urban explorer and random walk processes were run was obtained using the Python package `osmnx'. The urban explorer parameter $\beta$ was $1.5$, and the process was run for $T = 10^7$ timesteps, after which the distribution of $p_i$ was observed to be stationary.}
  \label{schematic}
\end{figure*}


The power of drive-by sensing hinges on the mobility patterns of the host fleet; wide coverage requires the  vehicles to densely explore a city's spatiotemporal profile. We call the extent to which a vehicle fleet achieves this their \textit{sensing power}. In what follows, we present a case study of the sensing power of taxi fleets.

Consider a fleet of sensor-equipped vehicles $\mathcal{V}$ moving through a city, sampling a reference quantity $X$ during a time period $\mathcal{T}$. We represent the city by a street network $S$, whose nodes represent possible passenger pickup and dropoff locations, and whose edges represent street segments potentially scannable by the vehicle fleet during $\mathcal{T}$. We use the proviso `potentially scannable', since some segments are never traversed by taxis in our data sets and so are permanently out of reach of taxi-based sensing, as further discussed in Supplementary Note 1. To model the taxis' movements we introduce the \textit{urban explorer} process, a schematic of which is presented in Figs.~\ref{schematic}(a)-(c). The model assumes that taxis travel to randomly chosen destinations via shortest paths, with ties between multiple shortest paths broken at random. Once a destination is reached, another destination is chosen, again at random, and the process repeats. To reflect heterogeneities in real passenger data, destinations in the urban explorer process are \textit{not} chosen uniformly at random. Instead, previously visited nodes are chosen preferentially: the probability $q_n$ of selecting a node $n$ is proportional to $1 + v_n^{\beta}$, where $v_n$ is the number of times node $n$ has been previously visited and $\beta$ is an adjustable parameter that depends on the city. This `preferential return' mechanism is known to capture the statistical properties of human mobility \cite{song2010modelling}, and as we show, also captures those of taxis.


To compare our model to data, we  quantify the \textit{sensing power of a vehicle fleet} as its covering fraction $\langle C \rangle$, defined as the average fraction of street segments in $S$ that are `covered' or sensed by a taxi during time period $\mathcal{T}$, assuming that $N_V$ vehicles are selected uniformly at random from the vehicle fleet $\mathcal{V}$. (In Supplementary Note 5 we consider an alternate definition.) 



We have computed $\langle C \rangle$ for 10 data sets from 9 cities: New York (confined to the borough of Manhattan), Chicago, Vienna, San Francisco, Singapore, Beijing, Changsha, Hangzhou, and Shanghai. (We used two independent data sets for Shanghai, one from 2014 and the other from 2015. For the 2015 data set, we chose the subset of taxi trips starting and ending in the subcity ``Yangpu'', and hereafter consider it a separate city.) Each data set consists of a set of taxi trips. The representation of these trips  differs, however, by city, and roughly falls into two categories. The Chinese cities comprise the first category, in which the GPS coordinates of each taxi's trajectory were recorded, along with the identification (ID) number of the taxi. Knowing taxi IDs lets us calculate $\langle C \rangle$ explicitly as a function of the number of sensor-equipped vehicles $N_V$, as desired. Accordingly, we call these the ``vehicle-level'' data sets. For the remaining cities, however, trips were recorded without taxi IDs; in these cases we know only how many trips were taken, not how many taxis were in operation for the duration of our data sets. (Although taxi IDs are available for Yangpu and New York City, for reasons discussed in Supplementary Note 1 we exclude them from the vehicle-level data sets). So for these ``trip-level'' data sets we can only calculate the dependence of $\langle C \rangle$ on $N_T$, the number of trips, which serves as an indirect measure of the sensing power. Finally, since we represent cities by their street networks, and not as domains in continuous space, we map GPS coordinates to street segments using OpenStreetMap, so that trips are expressed by sequences of street segments $(S_1, S_2, \dots)$. 

\begin{figure*}
 \includegraphics[width= 2\columnwidth]{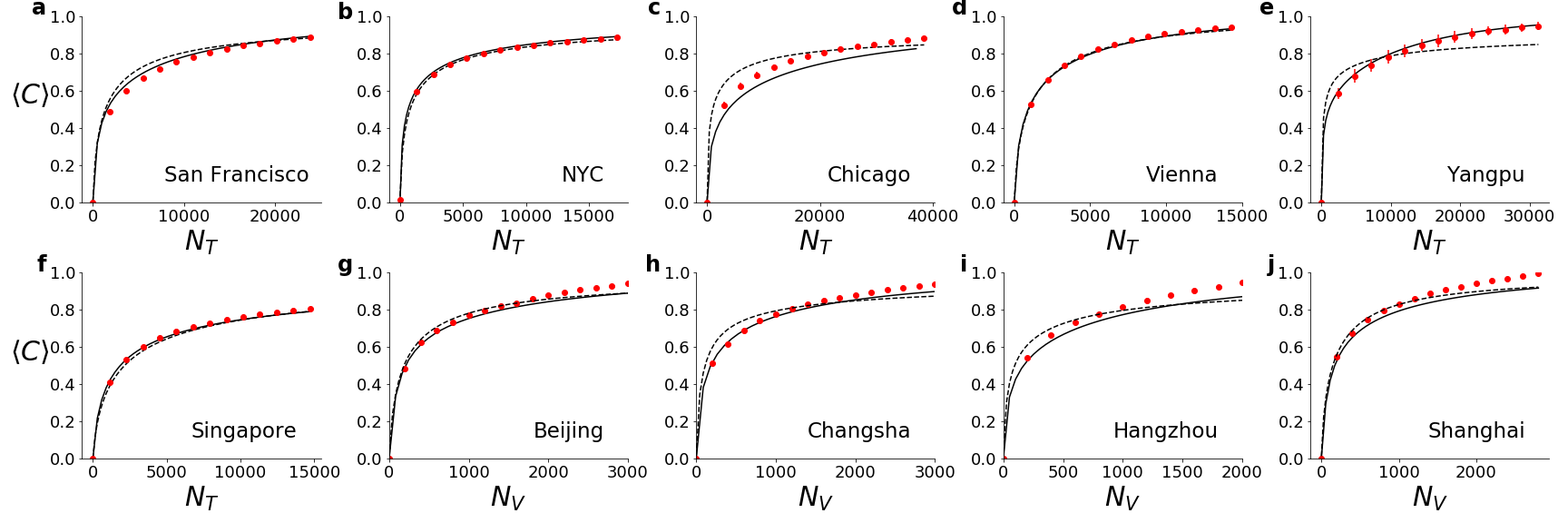}
 \caption{ \textbf{Sensing power $\mathbf{\langle C \rangle}$.} Theoretical and empirical street-covering fractions $\langle C \rangle $ for all data sets. Panels (a)-(f) show the trip-level data, where the dependent variable is the number of trips $N_T$, and (g)-(j) show the vehicle-level data, where the dependent variable is the number of vehicles $N_V$. Thick and dashed curves show the analytic predictions for $\langle C \rangle$ using $p_i$ estimated from data and the urban explorer process respectively. Red dots show the empirical $\langle C \rangle$, whose calculation we describe in Supplementary Note 2. Notice in (a)-(f) the number of trips needed to scan half a city's street segments, $N_T^*$, is remarkably low: $\sim 2000 = 10\%$, and in panels (g)-(j), $N_V^* \sim 5\% $. Exact figures for each $N_T^*, N_V^*$ are given in Supplementary Note 4. We list the city name, date, and parameter $\beta$ for each city below: (a) San Francisco, 05/24/08, $\beta = 0.25$ (b) New York City, 01/05/11, $\beta = 1.5$ (c) Chicago, 05/21/14, $\beta= 3.0$ (d) Vienna, 03/25/11, $\beta = 0.25$ (e) Yangpu, 04/02/15, $\beta = 2.75 $  (f) Singapore, 02/16/11, $\beta = 1.0$ (g) Beijing, 03/01/14, $\beta = 1.0$ (h) Changsha, 03/01/14, $\beta = 1.75$ (i) Hangzhou, 04/21/15, $\beta = 1.25$ (j) Shanghai, 03/06/14, $\beta = 0.75$.}
\label{C_all_cities_combined}
\end{figure*}

We find that, despite its simplicity, the urban explorer process captures the statistical properties of real taxis' movements. Specifically, it produces realistic distributions of \textit{segment popularities} $p_i$, the relative number of times each street segment is sensed by the fleet $\mathcal{V}$ during $\mathcal{T}$ (in turn, these $p_i$ allow us to calculate our main target, $\langle C \rangle$). Figure~\ref{schematic}(d) shows the empirical distribution of the $p_i$ obtained from our New York data set (brown histogram). The distribution is heavy tailed and follows  Zipf's law (this is also true of the other cities; see Supplementary Figure 2). The distribution predicted by the urban explorer process (blue histogram) is consistent with the data. This good agreement is surprising. One might expect the many factors absent from the urban explorer process -- variations in street segment lengths and driving speeds, taxi-taxi interactions, human routing decisions, heterogeneities in passenger pickup and dropoff times and locations -- would play a role in the statistical properties of real taxis. Yet our results show that, at the macroscopic level of segment popularity distributions, these complexities are unimportant. Moreover, the agreement of the model and the data is not trivial. Compare, for example, the predictions of a random walk model (Fig.~\ref{schematic}(e)). With their skewed unimodal distribution, the random walk $p_i$ fail to capture the qualitative behavior observed in the data.



Having obtained the segment popularities $p_i$, we can predict the sensing power $\langle C \rangle_{N_V}$ analytically by using a simple ball-in-bin model. We treat street segments as `bins' into which `balls' are placed when they are traversed by a sensor-equipped taxi. Using the segment popularities $p_i$ as the bin probabilities, we derive (see \textit{Methods}) the approximate expression

\begin{equation}
\langle C \rangle_{N_V} \approx 1 -  \frac{1}{N_S}\sum_{i=1}^{N_S} (1-p_i)^{\langle B \rangle*N_V}.
\label{final_C_mbar1_vehicle1}
\end{equation}
\noindent
Here $\langle B \rangle$ is the average distance (measured in segments) traveled by a taxi chosen randomly from $\mathcal{V}$ during $\mathcal{T}$. The `trip-level' expression $\langle C \rangle_{N_T}$ is the same as Eq.~\eqref{final_C_mbar1_vehicle1} with $\langle B \rangle$ replaced by $\langle L \rangle$, the average number of segments in a randomly selected trip. (See Methods, Eq.\eqref{final_C_mbar1_trip}.)

Figure~\ref{C_all_cities_combined} compares the analytic predictions for $\langle C \rangle$ against our data for a reference period of $\mathcal{T} = 1$ day (see Supplementary Note 2 for how the empirical $C$ were calculated). We tested the prediction \eqref{final_C_mbar1_vehicle1} in two ways: using $p_i$ estimated from our data sets (thick line), and using $p_i$ estimated from the stationary distribution of the urban explorer process (dashed line). In both cases theory agrees well with data, although the latter estimate is less accurate (as expected, it being derived from a model). Note that the $\langle C \rangle$ curves from different cities in Fig.~\ref{C_all_cities_combined} are strikingly similar. This similarity stems from the near-universal distributions of $p_i$ (shown in Supplementary Figure 1 and discussed in Supplementary Note 2) and suggests $\langle C \rangle$ might also be universal. 

\begin{figure}
  \includegraphics[width= \columnwidth]{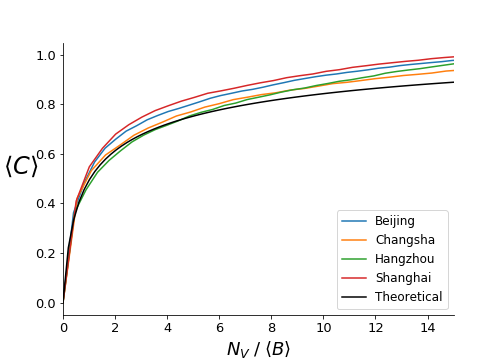}
  \caption{\textbf{Scaling collapse.} Empirical street-covering fractions $\langle C \rangle$ versus normalized number of sensor-equipped vehicles $N_{V} / \langle B \rangle $ from the four vehicle-level data sets. Remarkably, with no adjustable parameters, the curves for all four data sets fall close to the same curve, suggesting that at a statistical level, taxis cover street networks in a universal fashion. For each data set, the estimated values of $\langle C \rangle$ were found by drawing $N_V$ vehicles at random and computing the covering fractions. This process was repeated $10$ times. The variance in each realization was $O(10^{-3})$, so error bars were omitted. For the theoretical curve Eq.~\eqref{final_C_mbar1_vehicle1}, the $p_i$ were estimated using the urban explorer process with $\beta = 1.0$ on the Beijing street network. The choice of Beijing was arbitrary since, recall, the $p_i$ from different cities are nearly universal.}
  \label{C_all_cities_vehicles_scaled}
\end{figure}

Figure~\ref{C_all_cities_vehicles_scaled} tests for universality in the $\langle C \rangle$ curves. Using the vehicle-level data, we rescale $N_V \rightarrow N_V / \langle B \rangle$, which removes the city-dependent term $\langle B \rangle$. (We assume the $p_i$ are universal, so we do not rescale them.) With no other adjustments, the resulting curves nearly coincide, as if  collapsing on a single, universal curve. (The fidelity of the collapse however varies by day; see Supplementary Note 3). In Supplementary Figure 10 we perform the same rescaling for the trip-level data, which shows a poorer collapse. However since these data sets are of lower quality than the vehicle-level data, less trust should be placed in them. Hence, given the good collapse of the vehicle-level data, we conclude the sensing power of vehicle fleets, as encoded by $\langle C \rangle$, might be universal. 

The fast saturation of the $\langle C \rangle$ curves tells us taxi fleets have large, but limited, sensing power; popular street segments are easily covered, but unpopular segments, being visited so rarely, are progressively more difficult to reach. A law of diminishing return is at play, which means that while scanning an entire city is difficult, a significant fraction can be scanned with relative ease. In particular, as detailed in Supplementary Note 4, about $ 65 \%$ of vehicles are required to scan 80\% of a city's scannable street segments, but $50 \%$ of segments are covered by just $N_{V}^* \sim 200 \sim 5 \%$ of vehicles (and at the trip level $N_{T}^* \sim 2000 \sim 10 \% $). Most strikingly, as shown in Supplementary Figure 11, one third of the street segments in Manhattan are sensed by as few as ten random taxis!  In fact, because our estimates for $B$ are lower bounds (see Supplementary Note 2), the above quoted values for $N_V^*$ are likely lower. These remarkably small values of $N_V^*$ and $N_T^*$ are encouraging findings, and certify that drive-by sensing is readily feasible at the city scale, thus achieving the main goal of our work. 


There are many ways to extend our results. To keep things simple, we characterized the sensing power of taxi fleets with respect to the simplest possible cover metric: the raw number of segments traversed by a taxi at least once, $C= \sum 1_{(M_i \geq 1)}$ (where as defined in \textit{Methods}, $M_i$ is the number of times the $i$-th segment is sensed at the end of the reference period). A more general metric would be $C = \sum b_i 1_{(M_i \geq 1)}$, where $b_i$ could represent the length of the segment or an effective sensing area. Also for simplicity, we confined our analysis to the fixed reference period of a day. This restriction could be relaxed by describing the segment popularities $p_i$ by a time-dependent Poisson process with densities estimated from data.

Taxis traveling in cities share some of the features of non-standard diffusive processes. Like Levy walks \cite{shlesinger1986levy,blumen1989transport}, or the run-and-tumble motion of bacteria \cite{schnitzer1993theory}, their movements are partly regular and partly random. As such, they produce stationary densities on street networks that obey Zipf's law, contrary to a standard random walk. Future work could examine if other aspects of taxis' spreading behavior are also unusual. Perhaps the hybrid motion exemplified by taxis offers advantages in graph exploration \cite{tadic2003exploring}, foraging \cite{viswanathan2011physics}, and other classic applications of stochastic processes \cite{weiss1983random, sp_physics2}. 



The work most closely related to drive-by sensing is on `vehicle sensor networks' \cite{van2017quality}. Here, sensors capable of communicating with each other are fitted on vehicles, resulting in a dynamic network. The ability to share information enables more efficient, `cooperative' sensing, but has the drawback of large operational cost. Most studies of vehicle sensor networks are therefore in silico \cite{gerla2012vehicular}. Since the sensors used in drive-by sensing do not communicate, drive-by sensors are significantly cheaper to implement than vehicle sensor networks.


Vehicles other than taxis can be used for drive-by sensing. Candidates include private cars, trash trucks, or school buses. Since putting sensors on private cars might lead to privacy concerns, city-owned buses or trucks seem better choices for sensor hosts. The mobility patterns of school buses and trash trucks are however different to those of taxis; they follow fixed routes at fixed times, limiting their sensing power. The regularity in their motion opens up the possibility of `targeted sensing'. Should authorities want specific areas monitored at specific times, then sensors could be deployed on subsets of buses and trucks whose routes coincide with those sensing goals. This would yield more reliable coverage than that of taxis, whose random movements imply that sensing goals can only be probabilistically achieved.  The downside of targeting sensing is that the spatiotemporal volume defined by the scheduled routes of trucks and buses is small compared to that of taxis. Therefore for wider, more homogeneous cover, taxis are the better choice of sensor host.

The diverse data supplied by drive-by sensing have broad utility. High-resolution air-quality readings can help combat pollution, while measurements of air temperature and humidity can help improve the calibration of meteorological models \cite{mead2013use,katulski2010mobile} and are useful in the detection of gas leaks \cite{murvay2012survey}. 
Degraded road segments can be identified with accelerometer data, helping inform preventive repair \cite{nadeem2013mobile, wang2014framework}, while pedestrian density data can be helpful in the modeling of crowd dynamics \cite{kjaergaard2012mobile}. Finally, information on parking-spot occupancy, WiFi access points, and street-light infrastructure -- all obtainable with modern sensors -- will enable advanced city analytics as well as facilitate the development of new big data and internet-of-things services and applications.

In short, drive-by sensing will empower urban leaders with rich streams of useful data. Our study reveals these to be obtainable with remarkably small numbers of sensors. 





  \section*{Acknowledgments}
  The authors would like to thank Allianz,  Amsterdam Institute for Advanced Metropolitan Solutions, Brose, Cisco, Ericsson, Fraunhofer Institute, Liberty Mutual Institute, Kuwait-MIT Center for Natural Resources and the Environment, Shenzhen, Singapore- MIT Alliance for Research and Technology (SMART), UBER, Vitoria State Government, Volkswagen Group America, and all the members of the MIT Senseable City Lab Consortium for supporting this research. Research of S.H.S. was supported by NSF Grants DMS-1513179 and CCF-1522054.


\section*{Methods}
We wish derive an expression for the sensing power of a vehicle fleet. We quantify this by their covering fraction $\langle C \rangle_{N_V}$, the average fraction of street segments covered at least once when $N_V$ vehicles move on the street network $S$ according to the urban explorer process, during a reference period $\mathcal{T}$. Given the non-trivial topology of $S$ and the non-markovian nature of the urban explorer process, it is difficult to solve for $\langle C \rangle_{N_V}$ exactly. We can however derive a good approximation. It turns out that it is easier to first solve for the `trip-level' $\langle C \rangle_{N_T}$ metric, that is, when $N_T$, the number of trips in the dependent variable, so we begin with this case (the `vehicle-level' expression $\langle C \rangle_{N_V}$ then follows naturally).

Imagine we have a population $\mathcal{P}$ of taxi trajectories (defined, recall, as a sequence of street segments). The source of this population $\mathcal{P}$ is unimportant for now; it could come from a taxi (or fleet of taxis) moving according to the urban explorer process, or from empirical data, as we later discuss. Given $\mathcal{P}$, our strategy to find $\langle C \rangle_{N_T}$ is to map to a ``ball-in-bin process'': we imagine street segments as bins into which balls are added when they are traversed by a trajectory taken from $\mathcal{P}$. Note that, in contrast to the traditional ball-in-bin process, a random number of balls are added at each step, since taxis trajectories have random length.

\textbf{Trajectories with unit length}. Let $L$ be the random length of a trajectory. The special case of $L = 1$ is easily solved, because then drawing $N_{T}$ trips at random from $\mathcal{P}$ is equivalent to placing $N_B$ balls into $N_S$ bins, where $N_S$ is the number of segments, and each bin is selected with probability $p_i$. As indicated by the notation, we estimate these with the segment popularities discussed in the main text (we discuss this more later). Let $\vec{M} = (M_1, M_2, \dots, M_{N_S})$, where $M_i$ is the number of balls in the $i$-th bin. It is well known that the $M_i$ are multinomial random variables,
\begin{equation}
\vec{M} \sim \text{Multi}(N_{T}, \vec{p})
\end{equation}
where $\vec{p} = (p_1, p_2, \dots p_{N_S})$. The (random) fraction of segments covered is 
\begin{equation}
C = \frac{1}{N_S} \sum_{i=1}^{N_S} 1_{(M_i \geq 1)}
\end{equation}
where $1_A$ represents the indicator function of random event $A$. The expectation of this quantity is 
\begin{equation}
\langle C \rangle_{(N_T, L=1)} = \frac{1}{N_S} \sum_{i=1}^{N_S} \mathbb{P}_{N_T}(M_i \geq 1)
\label{tempC}
\end{equation}
\noindent
(note we introduce $L$ as a subscript for explanatory purposes). The number of balls in each bin is binomially distributed $M_i \sim Bi(N_B, p_i)$. The which has survival function $\mathbb{P}(M_i \geq 1) = (1 - (1-p_i)^{N_B})$. Substituting this into \eqref{tempC} gives the result
\begin{equation}
\langle C \rangle_{(N_B, L=1)} = 1-  \frac{1}{N_S} \sum_{i=1}^{N_S} (1-p_i)^{N_{B}}. \label{C_Lone}
\end{equation}

\textbf{Trajectories with fixed length}. Trajectories of fixed (i.e. non-random) length $L > 1$ impose \textit{spatial} correlations between the bins $M_i$ (recall that in the classic ball and bin problem, the $M_i$ are already correlated, since their sum is constant and equal to the total number of balls added $N_B$). This is because trajectories are contiguous in space; a trajectory that covers a given segment is more likely to cover neighboring segments. Given the non-trivial topology of the street network $S$, the correlations between bins are hard to characterize. To get around this, we make the strong assumption that for $N_T \gg 1 $ the spatial correlations between bins are asymptotically zero. This assumption greatly simplifies our analysis. It lets us re-imagine the ball-in-bin process so that adding a trajectory of length $L$ is equivalent to adding $L$ balls into \textit{non-contiguous} bins chosen randomly according to $p_i$. Then, selecting $N_T$ trajectories of length $L$ from $\mathcal{P}$ is equivalent to throwing $N_B = L*N_T$ balls into $N_S$ bins $\langle C \rangle_{(N_T,L_{fixed})} =  \langle C \rangle_{(N_T*L,L = 1)}$. Hence the expected coverage is a simple modification of \eqref{C_Lone}: 
\begin{equation}
\langle C \rangle_{(N_B, L=1)} = 1-  \frac{1}{N_S} \sum_{i=1}^{N_S} (1-p_i)^{ L* N_T}. \label{C_Lfixed}
\end{equation}
 

Assuming neighboring segments are spatially uncorrelated is a drastic simplification, and effectively removes the spatial dimension from our model. Yet surprisingly, as we will show, it leads to predictions that agree well with data. 



\textbf{Trajectories with random lengths}. Generalizing to random $L$ is straightforward. Let $S_{N_T} = \sum_{i=1}^{N_T} L_i$ be the number of segments covered by $N_T$ trajectories. By the law of total expectation 
\begin{equation}
\langle C \rangle_{(N_T,L)} = \sum_{n=0}^{\infty} \langle C \rangle_{(n,L_{fixed})}\mathbb{P}(S_{N_T} = n). \label{temp} 
\end{equation}
The first term in the summand is given by \eqref{C_Lfixed}. For the second term we need to know how the trajectory lengths are distributed. In Supplementary Figure 4 we show $L \sim Lognormal(\tilde{\mu}, \tilde{\sigma}^2) $. It is known that a sum of lognormal random variables is itself approximately lognormal $S_{N_T} \sim Lognormal(\mu_S, \sigma_S^2)$, for some $\mu_S$ and $\sigma_S$. There are many  different choices for $\mu_S, \sigma_S$; for a review see \cite{lognormal}. We follow the Fenton-Wilkinson method, in which $\sigma_S^2 = \ln{( \frac{\exp{{\tilde{\sigma}^2} -1}}{N_T} + 1 )}$ and $\mu_S = \ln{ (N_T \exp( \tilde{\mu}))} + (\tilde{\sigma}^2 - \sigma_S^2) / 2$. Then,
\begin{equation}
\mathbb{P}(S_{N_T} = n) = \frac{1}{ n \sigma_S \sqrt{2 \pi}} e^{ - \frac{ (\ln{n} - \mu_S)^2}{2 \sigma_S^2} }. 
\end{equation}
Substituting this into \eqref{temp} gives
\begin{equation}
\langle C \rangle_{(N_T, L)} = \frac{1}{N_S n \sigma_S \sqrt{2 \pi}} \sum_{n=0}^{\infty} \sum_{i=1}^{N_S} \big( 1 - (1-p_i)^n \big)  e^{ - \frac{ (\ln{n} - \mu_S)^2}{2 \sigma_S^2} }.
\label{semifinal_C}
\end{equation}
The above equation fully specifies the desired $\langle C \rangle_{(N_T,L)}$. It turns out however that the sum over $n$ is dominated by its expectation, so we collapse it, replacing $n$ by its expected value $\langle L \rangle*N_T$. This yields the much simpler expression $ \langle C \rangle_{(N_T,L)} =  \langle C \rangle_{(N_T*\langle L \rangle, L = 1)}$, or
\begin{equation}
\langle C \rangle_{N_T} \approx 1 -  \frac{1}{N_S} \sum_{i=1}^{N_S} (1-p_i)^{\langle L \rangle*N_T}
\label{final_C_mbar1_trip}
\end{equation}
which appears in the main text.

\textbf{Extension to vehicle level}. Translating our analysis to the level of vehicles is straightforward. Let $B$ be the random number of segments that a random vehicle in $\mathcal{V}$ covers in the reference period $\mathcal{T}$ (in Supplementary Figure 4 we show how $B$ are distributed in our data sets). Then we simply replace $\langle L \rangle$ with $\langle B \rangle$ in the expression for $\langle C \rangle_{N_T}$ to get $\langle C \rangle_{N_V}$,
\begin{equation}
\langle C \rangle_{N_V} \approx 1 -  \frac{1}{N_S}\sum_{i=1}^{N_S} (1-p_i)^{\langle B \rangle*N_V}.
\label{final_C_mbar1_vehicle}
\end{equation}
\noindent
\textbf{Model parameters}. The parameters $\langle L \rangle, \langle B \rangle$ in \eqref{final_C_mbar1_vehicle} as easily estimated from our data sets (see Supplementary Note 2). The bin probabilities $p_i$ are trickier. They have a clear definition in the ball-in-bin formalism, but in our model, the interpretation is not as clean; they represent the probability that a \textit{subunit} of a trajectory taken at random from $\mathcal{P}$ covers the $i$-th segment $S_i$. As mentioned above, we estimate these with the segment popularities, which we calculate in two ways: (i) deriving them directly from our data sets; or (ii) from the urban explorer process (recall these methods led to similar distributions of $p_i$ as shown in Fig.~\ref{schematic}(e)).

\bibliographystyle{naturemag}
\bibliography{refs.bib}


\end{document}


\title{Supplemental Materials}

\maketitle


\section*{Supplementary Figures}

\begin{figure}[H]  \centering
 \includegraphics[width= 0.6 \columnwidth]{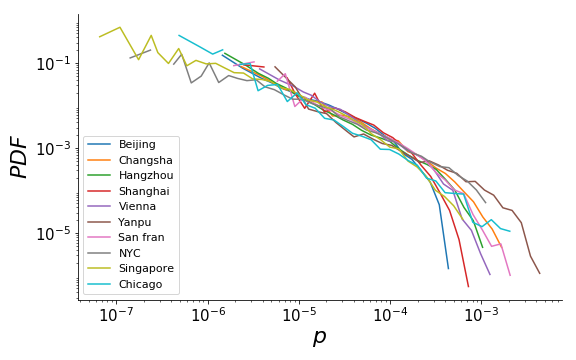}
 \caption{\textbf{Empirical segment popularities.} Log log plot of the distributions of segment popularities for each city. The curves are similar, but not universal as discussed in Supplementary Note 2. Note each curve has a non-straight tail, indicating a deviation from Zipf's law.}
\label{ps_all_cities}
\end{figure}

\begin{figure}[H]  \centering
 \includegraphics[width= 0.75 \columnwidth]{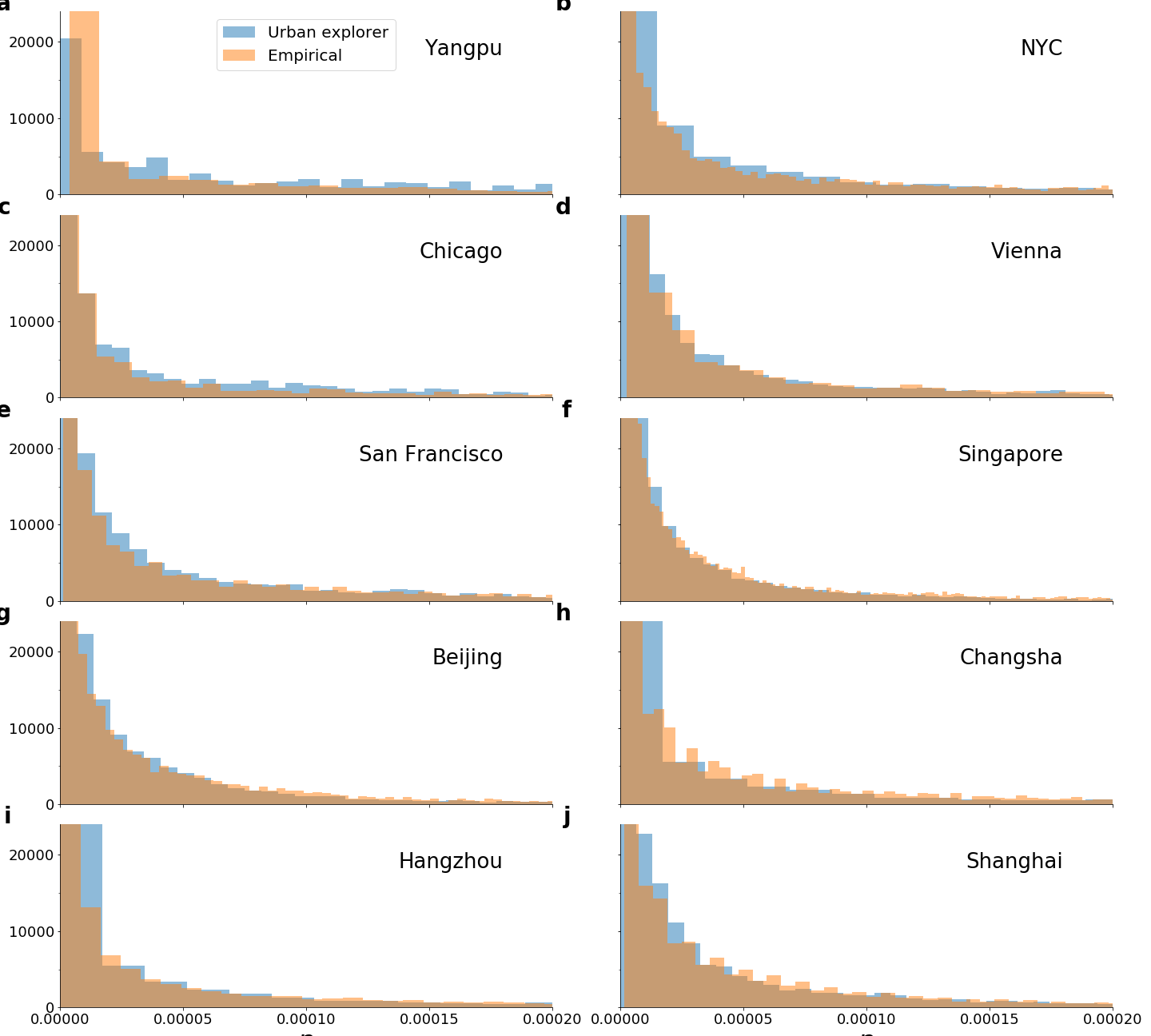}
 \caption{\textbf{Urban explorer segment popularities versus data.} Segment popularities $p_i$ derived from the urban explorer process (blue) and empirical data sets (orange) for all cities. $10^7$ timesteps were after which the distribution of $p_i$ were stationary. The bias parameters were (a) $\beta = 2.75$ (b) $\beta = 1.5$ (c) $\beta = 3.0$ (d) $\beta = 0.25$, (e) $\beta = 0.25$ (f) $\beta = 1.0$ (g) $\beta = 1.0$ (h) $\beta = 1.75$ (i) $\beta = 1.25$ (j) $\beta = 0.75$.}
\label{ue_versus_data}
\end{figure}



\clearpage

\begin{figure}\centering
  \includegraphics[width= 0.5 \linewidth]{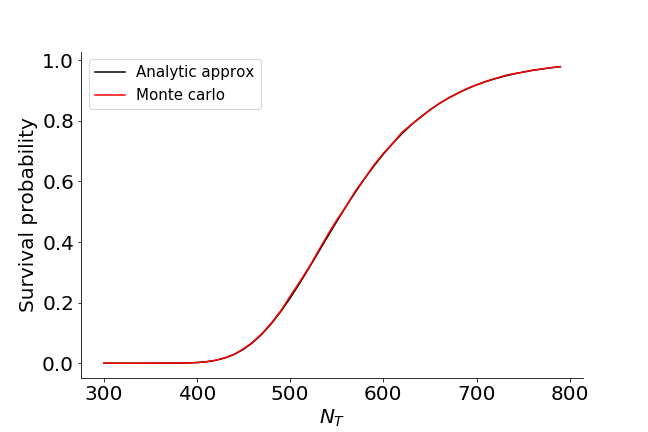}
  \caption{\textbf{Approximation of multinomial survival function}. Survival probability for multinomial distribution estimated from \eqref{final}, and via Monte Carlo. $10^5$ trials were used in each Monte Carlo approximation. 50 bins were used, with $p_i = 1/50$. The survival probability is defined as $\mathbb{P}(M_1 > b, M_2 > b, \dots)$. Here we took $b = 5$. Note the excellent agreement between theory and simulation (both curves lie on top of each other)}  \label{approx}
\end{figure}

\begin{figure} [H]  \centering
  \includegraphics[width= 0.75 \linewidth]{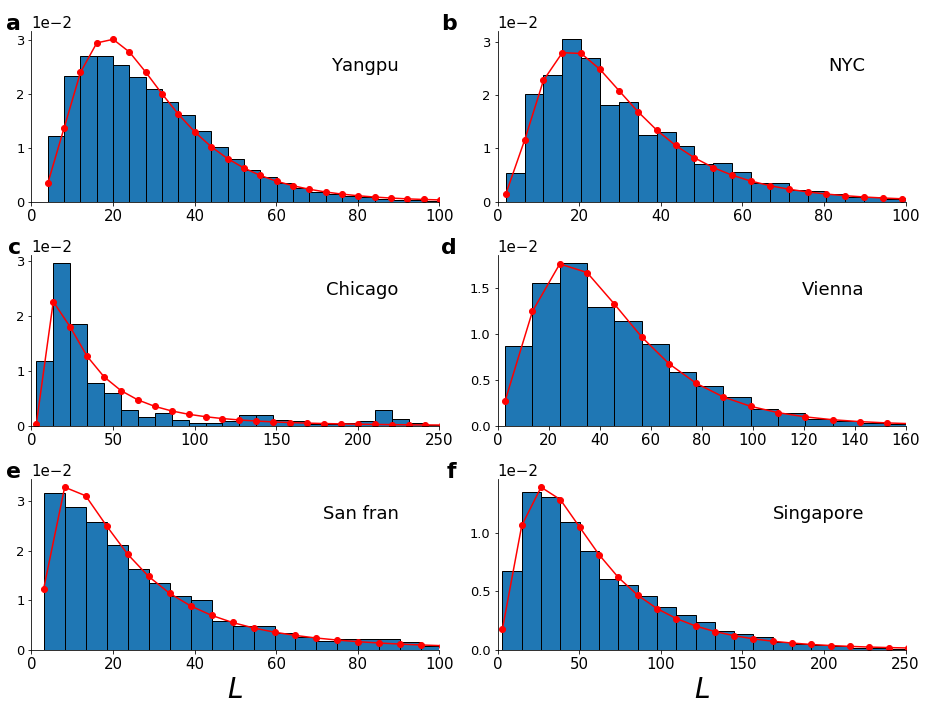}
  \caption{\textbf{Distributions of trajectories lengths for the trip-level datasets}. Histograms of the trajectory lengths $L$ during a given day for city. Red dotted lines show lognormal curves of best fit. We list the parameters of best fit $\mu, \sigma$, the sample mean $\langle L \rangle $, and the day the data were taken from for each subplot. Notice Chicago appears to have two humps. Data taken from other days are qualitatively similar. (a) Yangpu, 04/02/15, $(\mu, \sigma, \langle L \rangle) = (3.36, 0.52, 29.6)$ (b) NYC, 01/05/11, $(\mu, \sigma, \langle L \rangle) = (3.37, 0.57, 30.8)$ (c) Chicago, 05/21/14, $(\mu, \sigma, \langle L \rangle) = (3.36, 0.98, 51.02)$ (d) Vienna, 03/25/11, $(\mu, \sigma, \langle L \rangle) = (3.91, 0.51, 45.54)$ (e) San Fransisco, 05/24/08, $(\mu, \sigma, \langle L \rangle) = (2.99, 0.87, 28.90)$ (f) Singapore, 02/16/11, $(\mu, \sigma, \langle L \rangle) = (3.97, 0.68, 60.78)$.}
  \label{daily_L}
\end{figure}

\begin{figure} [H]  \centering
  \includegraphics[width= 0.75 \linewidth]{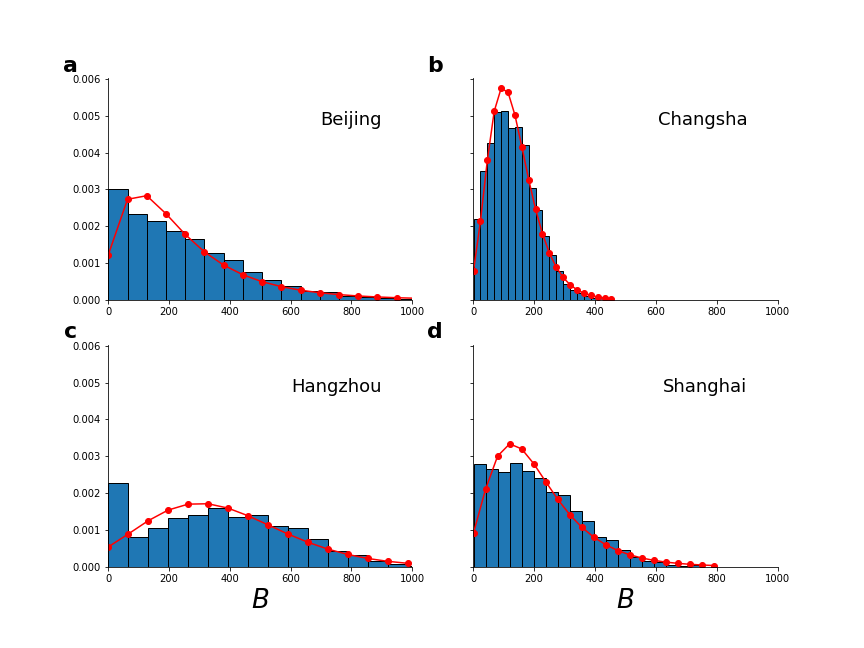}
  \caption{\textbf{Distributions of distance traveled by taxis for vehicle-level datasets}. Histograms of $B$, the distance traveled (measured in segments) by a taxi in a day for each city. Red dotted lines show lognormal curves of best fit. We list the parameters of best fit $\mu, \sigma$, the sample mean $\langle B \rangle $, and the day the data were taken from for each subplot (a) Beijing 03/02/13, $(\mu, \sigma, \langle B \rangle) = (5.56, 0.65, 245)$ (b) Changsha 03/02/14 $(\mu, \sigma, \langle B \rangle) = (5.46, 0.31, 131)$ (c) Hangzshou 04/22/14 $(\mu, \sigma, \langle B \rangle) = (5.13, 0.18, 366)$ (d) Shanghai 03/02/14 $(\mu, \sigma, \langle B \rangle) = (5.41, 0.35, 270)$.}
  \label{daily_L_vehicles}
\end{figure}

\begin{figure}[H]  \centering
 \includegraphics[width= 0.75 \columnwidth]{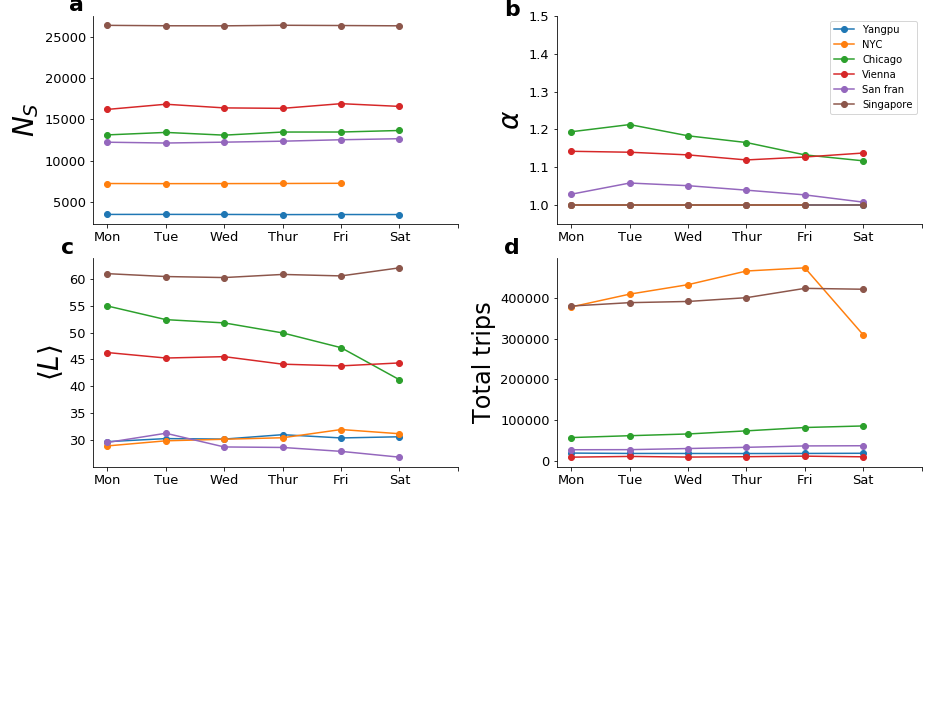}
 \caption{\textbf{Temporal fluctuations of trip-level datasets}. Generally speaking little daily variation in each quantity. (a) Number of scannable street segments (b) Best fit exponent in truncated power law $\alpha$. (c) Average  length of trajectory (d) Total number of trips.}
\label{temp_fluct}
\end{figure}

\begin{figure}[H]  \centering
 \includegraphics[width= 0.70 \columnwidth]{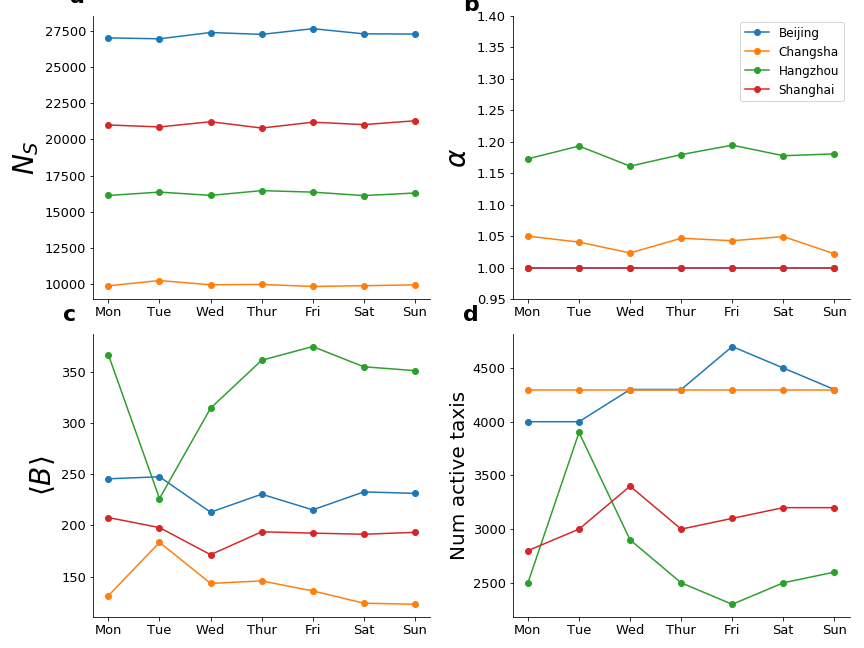}
 \caption{\textbf{Temporal fluctuations of vehicle-data}. Generally speaking little daily variation in each quantity. (a) Number of scannable street segments (b) Best fit exponent in truncated power law $\alpha$. (c) Average daily distance traveled by a taxi (d) Total number of trips.}
\label{temp_fluct_vehicles}
\end{figure}


\begin{figure} [H] \centering
  \includegraphics[width= 0.75 \columnwidth]{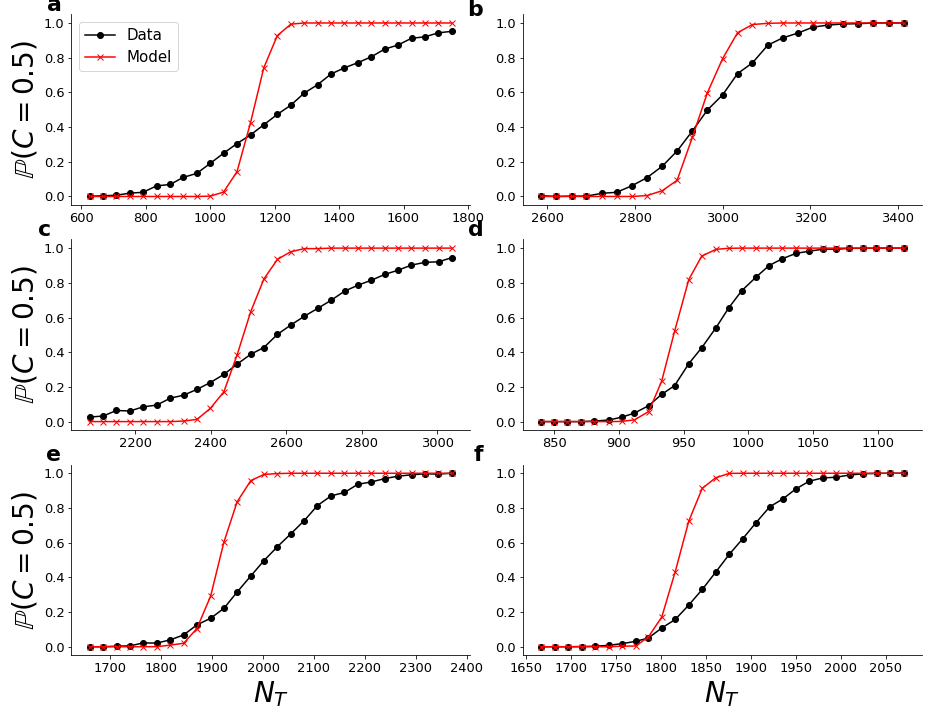}
  \caption{\textbf{Minimum street sampling problem}. Analytic prediction versus trip-level data. The red curve shows theoretical results, while the black curve shows probabilities estimated from data. The parameters for each subplot were $\bar{C} = 0.5$, $m = 1$. The number of trials used in the Monte Carlo estimate of $\mathbb{P}(C)$ was 1000. (a) Yangpu on  04/02/15 (b) NYC on 01/05/11 (c) Chicago on 05/21/14 (d) Vienna on 03/25/11 (e) San Fransisco on  05/24/08 (f) Singapore on 02/16/11}
  \label{P_all_cities}
\end{figure}

\begin{figure} [H] \centering
  \includegraphics[width= 0.75 \columnwidth]{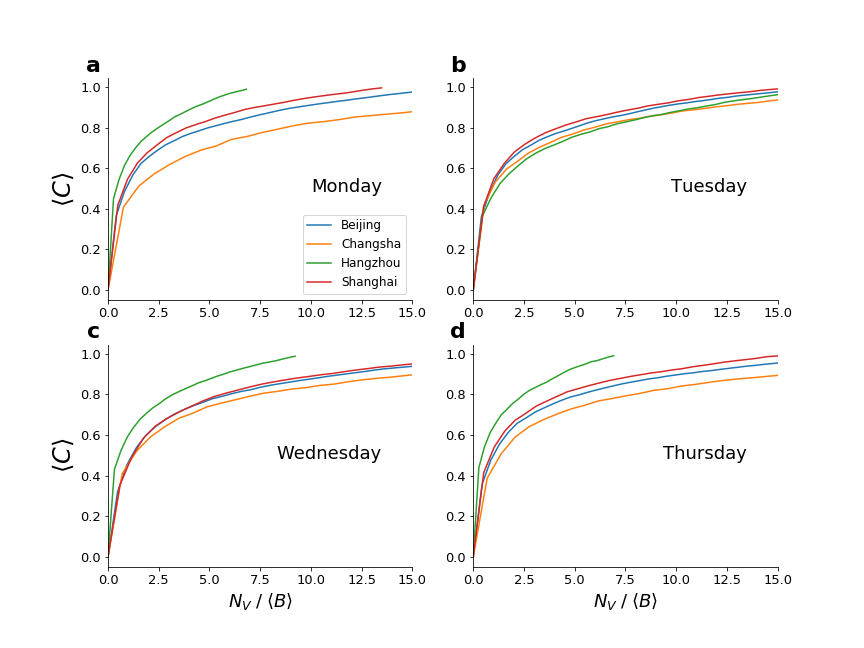}
  \caption{\textbf{Scaling collapse of vehicle-level data on different days} Counterpart of Figure 3 in the main text. As can be seen, a close approximation to a true scaling collapse is achieved only on Tuesday. Note the Hangzhou dataset has strong variations. This is not surprising, since as shown in Figure~\ref{temp_fluct_vehicles}, this dataset has strong temporal variations. In particular, $\langle B \rangle$ varies much more than the other datasets.}
  \label{scaling_collapse_vehicles}
\end{figure}

\begin{figure} [H] \centering
  \includegraphics[width= 0.6 \columnwidth]{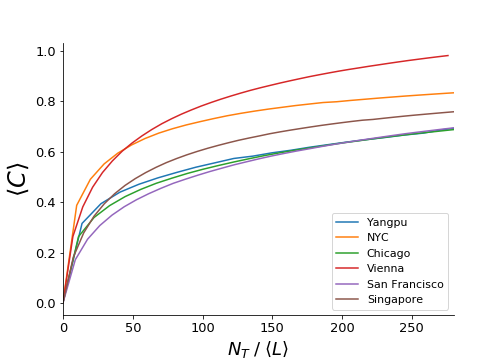}
  \caption{\textbf{Scaling collapse of trip-level data}. In contrast to vehicle-level datasets --  Supplementary Figure~\ref{scaling_collapse_vehicles} -- the trip-level datasets do not show universal behavior. There are however some trends. As can be seen the Chicago, San Francsico, and Yangpu datasets collapse to a common curve, where the other datasets do not. The data for each city are the same as those used in Figure 2 (main text). Trip data on different days show the same trends.}
  \label{scaling_collapse_trips}
\end{figure}

\begin{figure} [H] \centering
  \includegraphics[width= 0.6 \columnwidth]{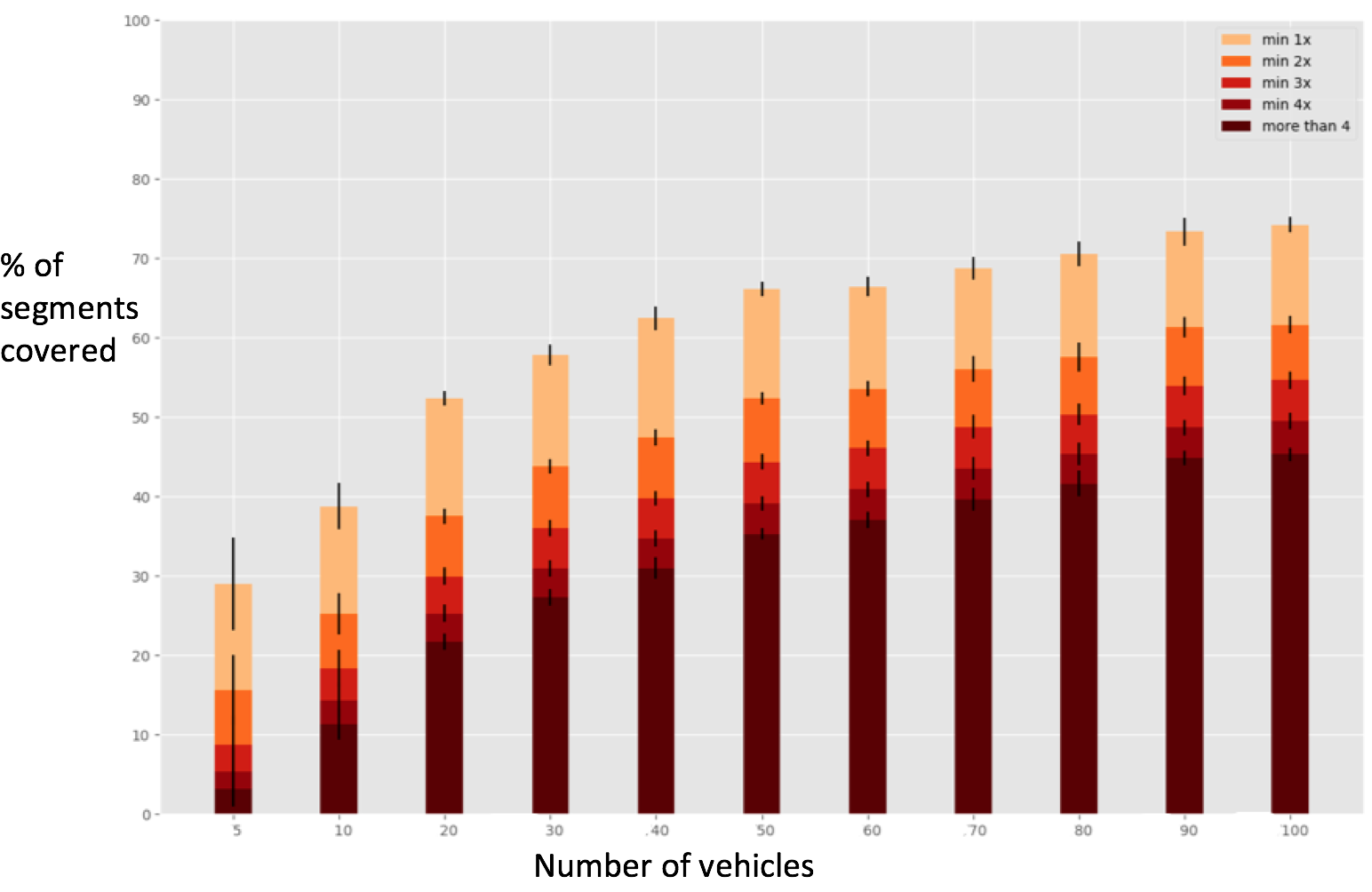}
  \caption{\textbf{Average segment coverage versus number of sensor-equipped taxis in Manhattan on 03/08/2011}. Different colors show results for different scanning thresholds. That is, the \% of segment at least $m$ times, where $m = 1,2,3,4$. Black lines show one standard deviation away from mean value. Notice that just $10$ vehicles scan more than a third of segments, while 30 scan more than half.}
  \label{manhattan_segments}
\end{figure}


\section*{Supplementary Note 1}

\textbf{Data sets}.
We have 10 real-world data sets from 9 cities: New York, Chicago, Vienna, San Francisco, Singapore, Beijing, Changsha, Hangszhou, and Shanghai. We had two independent data sets for Shanghai, independent in the sense they occurred on different years (2014 and 2015). For 2015 dataset, we selected only those trips starting and ending in the subcity ''Yangpu'', and hereafter consider it a separate city. The datasets were collected from various sources. Those from Beijing, Changsha, and, Hangszhou were provided by a third-party organization that collected driving data from taxi operation companies. The Shanghai datasets were provided by the `'1st Shanghai Open Data Apps 2015" (an annual competition).The New York dataset has been obtained from the New York Taxi and Limousine Commission for the year 2011 via a Freedom of Information Act request. The Vienna and Singapore datasets were provided to the MIT SENSEable City Lab by AIT and the Singapore government, respectively. The San Francisco and Chicago data sets were publicly available \cite{sanfran_dataset}, \cite{chicago_dataset}. Note the NYC, Vienna, San Francisco, and Singapore datasets were the same as used in previous studies \cite{santi2014}, \cite{scaling_law}. 

The four data sets from Chinese cities were very large ($\sim$ GB worth of data per day). For computational convenience, we therefore subsampled the datasets by selecting only those trips which occurred in a $20$ km box surrounding the city center. The city center was found using OpenStreetMap, the GPS coordinates of which were $(39.9059631, 116.3912480)$, $(28.1979483, 112.9713300)$, $(30.2489634, 120.2052342)$, $(31.2253441, 121.4888922)$ respectively.

The temporal range of the data sets was not uniform. NYC was the most comprehensive, consisting of a years worth of taxi trips in Manhattan. The remaining data sets were for one week. The sizes of the cities was also different. We show this in Figures~\ref{temp_fluct}(a) and~\ref{temp_fluct_vehicles}(a) by showing $N_S$, the number of scannable segments, for each city over the course of a week.

\begin{table}[]
\centering
\label{data_table}
\caption{Properties of data sets}
\label{data}
\begin{tabular}{|p{2.2 cm}|p{2.0cm}|p{1.5 cm}|p{5.0cm}|p{1.1 cm}|p{0.9 cm}|p{1.0 cm}|}
\hline
        \textbf{City}  & \textbf{Trajectories} & \textbf{Taxi Ids}  & \textbf{Temporal range} & $N_{S,total}$ & $N_S$ & $\frac{N_{S,total}}{N_S}$
\\[0.15 cm] \hline
Yangpu  & Real (GPS)            & Yes              & 1 Week: \; 04/01/15 -- 04/04/15  & 2919 & 2657 & 0.94  \\
NYC  & Generated             & Yes              & 1 Year: \; 12/31/10 -- 12/31/11 & 7954 & 7265 & 0.91\\
Chicago   & Generated             & No              & 1 Week: \; 06/23/14 -- 06/30/14 & 24054 & 12492  & 0.52 \\ 
Vienna    & Generated             & No               & 1 Week: \; 03/07/11 -- 10/07/11 & 24054 & 15775 & 0.66 \\
San Francisco  & Generated             & No               & 1 Week: \; 05/21/08 -- 05/28/08 & 15453 & 11708 & 0.76 \\
Singapore & Generated             & No               & 1 Week: \; 02/21/11 -- 02/28/11 & 32362 & 25255 & 0.78 \\
Beijing  & Real (GPS)            & Yes              & 1 Week: \; 03/01/14 -- 03/07/14 & 54665  & 27024 & 0.49  \\
Changsha  & Real (GPS)            & Yes              & 1 Week: \; 03/01/14 -- 03/07/14 & 18067 & 9882 & 0.55  \\
Hangszhou  & Real (GPS)            & Yes              & 1 Week: \; 04/21/15 -- 04/28/15 & 39056  & 16125.0 & 0.41 \\
Shanghai  & Real (GPS)            & Yes              & 1 Week: \; 03/01/14 -- 03/07/14 & 49899  & 21002 & 0.49
\\[0.1 cm] \hline
\end{tabular}
\end{table}

Each data set consists of a set of taxis trips. The representation of these trips differs by data set. For the Chinese cities, a trip is the set of GPS coordinates of the taxis position as its serves its passenger. Since in our model we represent cities by a street \textit{networks}, we convert the set of GPS coordinates to a \textit{trajectory}. (Recall in the main text we defined a trajectory as a sequence of street segments $T_r = (S_{i_1}, S_{i_2}, \dots)$). We matched the taxi trajectories to OpenStreetMap (driving networks) following the idea proposed in \cite{GPS_convert}, which using a Hidden Markov Model to find the most likely road path given a sequence of GPS points. The HMM algorithm overcomes the potential mistakes raised by nearest road matching, and is more robust when GPS points are sparse.

For the remaining data sets, each trip $i$ is represented by a GPS coordinate of pickup location $O_i$ and dropoff location $D_i$ (as well as the pickup times and dropoff times). As for the Chinese cities, we snap these GPS coordinates to the nearest street segments using OpenStreetMap. We do not however have details on the trajectory of each taxi (the intermediary path taken by the taxi when brining the passenger from $O_i$ to $D_i$.) We thus needed to approximate trajectories. We had two methods for this, one sophisticated, one simple. The sophisticated method was for the Manhattan dataset. Here, as was done in \cite{santi2014}, we used hour-by-hour variability in the traffic congestion, we did X. For the remaining cities, we used the simple method of finding the weighted shortest path between $O_i$ and $D_i$ (where segments were weighted by their length). As we will show, in spite of the different representations of trajectories, the sensing properties of the taxi fleets from each city are very similar. This gives us confidence in X. 

Lastly, for five of the nine cities -- the Chinese cities plus NYC --- taxi trips are recorded with the ID of the taxi which completed that trip. Hence for these `vehicle-level' datasets we can calculate $\langle C \rangle_{N_V}$ -- the sensing potential of a fleet as a function of the number of constituent vehicles $N_V$ directly. For the remaining cities, it is unknown which taxis completed which trips. Hence for these 'trip-level' data sets, we can solve only for $\langle C \rangle_{N_T}$. Hence we hereafter divide our datasets into these two categories -- `vehicle-level' and 'trip-level' -- and use these terms throughout the paper. For the sake of comparison, we decided to consider NYC and Yangpu part of the trip-level datasets. That way, the three different representation of trajectories feature in the trip levels datasets, giving more confidence in their results. 

We summarize all the properties of dataset discussed above in Supplementary Table 1 

In the main text, we noted that we analyzed subsets of the real-world street networks, those sub-networks containing only 'potentially scannable' edges. We had two reasons for doing this. First, there are some streets which are never traversed by taxis in our data sets; since there are permanently out of reach of taxi-based sensing, we do not consider them. Second, since city borders aren't well defined, the total number of streets in a city $N_{S, total}$ is also ill-defined (potential exceptions being Manhattan and Singapore, which have sharp borders). In light of these two complications, we consider only those streets which were traversed at least once by taxis in our dataset $N_S$, which in general is different to $N_{S,total}$.


\section*{Supplementary Note 2}
\textbf{Estimation of parameters from data sets.} 

There are three parameters in our model: $p_i$, the segment popularities, $B$, the random distance traveled by a taxi randomly selected from $\mathcal{V}$, and $L$, the random length of a trip (recall $B$ is needed for the vehicle-level data for which $\langle C \rangle$ is a function of $N_V$, the number of vehicles, and $L$ is needed for trip-level data, for which  $\langle C \rangle$ function of $N_T$.) Supplementary Figures~\ref{daily_L} and~\ref{daily_L_vehicles} shows the distributions $\mathbb{P}(L)$ and $\mathbb{P}(B)$ for each city on a given day. Note that while $L$, the random length of a trajectory (measured in segments), can be directly found from our datasets, $B$, the total distance traveled (again measured in segments) by a taxi during the reference period $T$ is not. This is because our data sets contain taxi \textit{trips} only  -- a trip implying a passenger is on board -- and do not include the distance traveled by taxis when they are empty. Hence, estimating $B$ from our datasets constitutes a \textit{lower} bound for the true $B$. 

Coming back to Supplementary Figures~\ref{daily_L} and~\ref{daily_L_vehicles}, we see the distribution $\mathbb{P}(L), \mathbb{P}(B)$ are well fit by lognormals (shown as red curves in the figures). The lognormal fits well in all cases, with exceptions being Chicago, and to a lesser extent, San Francisco (which is contrast to the others appears to be monotonically decreasing). In panel (c) of Supplementary Figures~\ref{temp_fluct} and~\ref{temp_fluct_vehicles} we show how $\langle L \rangle$ varies by day. There is little variation. In the other panels of these two figures, we show how the number of scannable street segments $N_S$, the total number of trips, the number of active taxis, as well as $\alpha$ (a parameter characterizing the distribution of the segment popularities $p_i$ -- we will discuss this parameter shortly) vary by day. In most cases, there is also little variation. These are encouraging findings, since they indicate the behavior of our model is general, not giving (significantly) different results on different days of the week.

To test the universalities in $p_i$, we fit each dataset to various heavy tailed distributions, listed in equation~\eqref{heavy_tailed}.

\begin{align}
 P_{exponential}(x) &= \lambda e^{- \lambda (x - x_{min})} \nonumber \\
 P_{power \; law}(x) &= (\alpha-1) x_{min}^{\alpha-1} x^{-\alpha} \nonumber \\
 P_{log \; normal}(x) &= x^{-1} \exp(-\frac{(\log x - \mu)^2}{2 \sigma^2} ) \nonumber  \\
 P_{stretched \; exponential}(x) &= \beta \lambda x^{\beta -1} e^{-\lambda (x^{\beta}- x_{min}^{\beta})} \nonumber \\
P_{truncated \; power law}(x) &= \frac{\lambda^{1-\alpha}}{\Gamma(1-\alpha,\lambda x_{min})} x^{-\alpha} e^{- \lambda x} \label{heavy_tailed}
\end{align}

We performed the fitting using the python package `powerlaw'. By default this package determines a minimum value $p_{min}$ below which data are discarded. Since we want to model the full $\mathbb{P}(p)$ (and not just the tail), we set this equal to the minimum value in our datasets. We show the results of the fittings in Table 2. For each city, either a truncated power laws or stretched exponentials was selected as the distribution which fit the data best. Thus, we only report the parameters for those two distributions (the parameters are defined by equations~\eqref{heavy_tailed}). As detailed in documentation of `powerlaw', parameters of best fit  are found by maximum liklihood estimation. We estimated errors in these parameters by bootstrapping: new data sets $(p_i^*)_{i^*=1}^{N_S}$ were drawn uniformly at random from the original data set $(p_i)_{i=1}^{N_s}$ 1000 times, best fit parameters were found for each of these 1000 realization, the standard deviation of which was taken as the standard error in each parameter. The `goodness of fit' measure for each distribution is quantified by the KS (kolmogorov-smirnoff) parameter $D$, defined by 
\begin{equation}
D = \max_x \Big| CDF_{empirical}(x) - CDF_{theoretical}(x) \Big|
\end{equation}
\noindent
where smaller $D$ values indicate better fits, and where $CDF$ denotes the cumulative density function. Finally, the liklihood-ratio test was used to compare the distribution of one fit to another. This has two parameters $\Lambda, p_1$. The sign of $\Lambda$ tells which distribution is more likely to have generated the data (positive means the first, negative means the second), while the $p_1$-value gives a measure of the confidence in the value of $\Lambda$ (the smaller, the more confident). We adopt the convention that $\Lambda > 0$ indicates the stretched exponential is preferred over the truncated power law (and $\Lambda < 0$ indicates the opposite).

As can be seen in Supplementary Table 2, the tests tell us $\mathbb{P}(p)$ of three of cities are best modeled by stretched exponentials, while the others are best modeled by truncated power laws. The values for $p_i$ were all $< O(10^{-26})$ (and as small as $O(10^{-222})$), so we truncated all values to zero. There are some mild similarities in the best fit parameters, but no evidence of a convincing trend. Hence, we conclude that the segment popularity distributions $\mathbb{P}(p)$ are not universal.

Like $\mathbb{P}(L)$ and $\mathbb{P}(B)$, there is little daily variation in $\mathbb{P}(p)$. We demonstrate this in Supplementary Figure~\ref{temp_fluct}(b) and~\ref{temp_fluct_vehicles}(b) where we show the maximum liklihood exponent $\alpha$ of the truncated power law fit measured day-by-day (for clarity, we do not display the $\beta$ parameter of the stretched exponential, but they show the same trends). 

\begin{widetext}[h!]
\vspace{1cm}
\centering
\begin{tabular}{ |P{2.5cm}|P{6cm}|P{5 cm}|P{2 cm}|}
 \multicolumn{4}{c}{\textbf{Maximum liklihood parameters}} \\
 \hline
   &  $(\lambda, \beta, D)_{\text{ \color{red}{stretched \; exponential}}}$ & $(\lambda, \alpha, D)_{\text{trunc. power law}}$ & $(\Lambda,p_1)$ \\
 \hline
 
Yangpu & $\Big( (1.3 \pm 0.1)*10^{6}, 0.266 \pm 0.003, 0.08 \Big)$ & $(5830 \pm 10, 1.132 \pm 0.004, 0.07)$ & $(-1327,0)$  \\
 
NYC &  $\Big((15 \pm 4)*10^{3}, 0.499 \pm 0.005, 0.03 \Big)$ & $(780 \pm 20, 1.00 \pm 10^{-6},0.25)$ & $(1600,0)$ \\

Singapore &  $\Big( (591 \pm 8)*10^{3}, 0.499 \pm 0.004, 0.02 \Big)$ & $(3400 \pm 200, 1 \pm (6*10^{-8}), 0.2)$ & $(3282,0)$ \\

Chicago &  $\Big( (3.4 \pm 0.9)*10^{6}, 0.187 \pm 0.005, 0.04 \Big)$ & $(650 \pm 30, 1.170 \pm 0.006, 0.03)$ & $(-208,0)$ \\

San Francisco &  $\Big( (47 \pm 7)*10^{5},0.257 \pm 0.005, 0.03 \Big)$ & $(1330 \pm 50, 1.156 \pm 0.008, 0.04)$ & $(-218,0)$ \\

Vienna &  $\Big( (41 \pm 0.5)*10^{5}, 0.293 \pm 0.006, 0.04 \Big)$ & $(2420 \pm 80, 1.196 \pm 0.008, 0.05)$ & $(-278,0)$ \\

Beijing &  $\Big( (1.2 \pm 0.4)*10^{5}, 0.42 \pm 0.002, 0.06 \Big)$ & $(5940 \pm 10, 1.00 \pm 10^{-6}, 0.08)$ & $(824,0)$ \\

Changsha &  $\Big( (7.5 \pm 0.2)*10^{5}, 0.34 \pm 0.003, 0.04 \Big)$ & $(1750 \pm 10, 1.02 \pm 0.02, 0.04)$ & $(248,0)$ \\

Hangzhou &  $\Big( (1.3 \pm 0.2)*10^{6}, 0.23 \pm 0.003, 0.05 \Big)$ & $(1770 \pm 20, 1.16 \pm 0.004, 0.04)$ & $(560,0)$ \\

Shanghai &  $\Big( (7.8 \pm 0.4)*10^{5}, 0.43 \pm 0.004, 0.05 \Big)$ & $(4970 \pm 10, 1.00 \pm 10^{-6}, 0.06)$ & $(564,0)$ \\

\hline 
 \label{MLE_estimates}
\end{tabular}
\vspace{1cm}
\end{widetext}

\textbf{Compare $C_{model}$ and $C_{data}$}. In the main text we compare our expression for $\langle C \rangle$ against data for a given reference period of a day. The empirical $\langle C \rangle$ were found by subsampling the datasets on a given day; random subsets were drawn from a day's worth of trips, and the average fraction of segments covered by those subsets was computed. As mentioned in the main text, we tested the analytic prediction two ways: using $p_i$ estimated by the stationary distributions of the urban explorer process (dashed line), and also directly from our datasets (thick line). In the latter case we calculated the distribution of $p_i$ for each day of the week (excluding Sunday), then used those to calculate six separate $\langle C \rangle$, the average of which is shown. This way, both temporal fluctuations and the bias of using the same datasets to estimate $p_i$ and the empirical $\langle C \rangle$ (which recall was calculated for a \textit{single} day) were minimized. We discuss this in more detail in Supplementary Note 2. For both these cases, the parameter $\langle B \rangle$ was estimated from datasets. In Supplementary Figures 4 and 5 we show the empirical distributions of $p_i$, $B$, and $L$, and show in Supplementary Figures 6 and 7 that they do not vary significantly on different days of the week.

\section*{Supplementary Note 3}
\textbf{Scaling Collapse}. We first discuss the vehicle-level data. In the main text we derived
\begin{equation}
\langle C \rangle_{(N_V)} = 1 -  \frac{1}{N_S}\sum_{i=1}^{N_S} (1-p_i)^{\langle B \rangle*N_V}.
\label{temp}
\end{equation}
which contains the parameters $p_i, \langle B \rangle$ and  $N_S$. Since $p_i$ and $N_S$ specify the distribution of $P(p_i)$, and since the $\mathbb{P}(p)$ are approximately universal across cities (see Supplementary Figure~\ref{ps_all_cities}), we only need to remove the parameter $\langle B \rangle$ from \eqref{temp} to make it city independent. Thus the simple rescaling $N_V \rightarrow N_V / \langle B \rangle$ gives the city-independent quantity
\begin{equation}
\langle C \rangle_{(N_V / \langle B \rangle)} = 1 -  \frac{1}{N_S}\sum_{i=1}^{N_S} (1-p_i)^{N_V}.
\end{equation}
We plot this in Supplementary Figure~\ref{scaling_collapse_vehicles} for different days. As can be seen, the quality of the data collapse varies by day. Hangzhou varies the most, which is to be expected, since it is this dataset set which has the highest temporal variation, as shown in Supplementary Figure~\ref{temp_fluct_vehicles}. 

We apply the same procedure to the trip-level data, except now the scaling is $N_T \rightarrow N_T / \langle L \rangle$. Figure~\ref{scaling_collapse_trips} shows the result. A universal scaling collapse is absent, although there is some similarities between the data sets; Chicago, Yangpu, and San Francisco and nearly coincident. The lack of full universal behavior is perhaps due to the inferior quality of the trip-level datasets (inferior because the trajectories are inferred).


\section*{Supplementary Note 4}

We here give explicit values for $N_T^*$ and $N_V^*$ the numbers of trips and vehicles needed to cover half of the city's scannable street segments, i.e. the solutions to $\langle C \rangle(N_T^*) = 0.5$ and $\langle C \rangle(N_V^*) = 0.5$.

\begin{table}[h!]
\centering
\label{data_table_Nstart}
\caption{Coverage statistics. $N_{T,total}$ refers to the total number of trips occurring on the specified day.}
\label{N_star_data}
\begin{tabular}{|p{1.5 cm}|p{1.5cm}|p{1.5 cm}|P{1.75cm}|P{1.5cm}|}
\hline
        \textbf{City}  & $N_T^*$ & $N_{T,total}$ & $N_T^* / N_{T,total}$ & Date
\\[0.15 cm] \hline
Yangpu  &  947           & 17571               & 5.40 \%                 &  04/02/15                 \\
New York  & 1179             & 466237               &  0.25 \%                  & 01/05/11                  \\
Chicago   & 2619             & 67848               &  3.86 \%                  & 05/21/14                  \\
Vienna    & 1010             & 10948                & 9.23 \%                  & 03/25/11                  \\
San fran  & 1923             & 36089                &  5.33 \%                 & 05/24/08                  \\
Singapore & 1782             & 401879                &  0.44 \%                 & 02/16/11
\\[0.1 cm] \hline
\end{tabular}
\end{table}

\begin{table}[h!]
\centering
\label{data_table_Nstart}
\caption{Coverage statistics. $N_{V,total}$ refers to the total number of taxis on the specified day.}
\label{N_star_data}
\begin{tabular}{|p{1.5 cm}|p{1.5cm}|p{1.5 cm}|P{1.75cm}|P{1.5cm}|}
\hline
        \textbf{City}  & $N_V^*$ & $N_{V,total}$ & $N_V^* / N_{V,total}$ & Date
\\[0.15 cm] \hline
Beijing & 211             & 4000                &  5.28 \%                 & 03/01/14 \\
Changsha & 227             & 4300                &  5.28 \%                 & 03/01/14  \\
Hangzhou & 132             & 2500                &   5.28 \%                 & 04/21/15 \\
Shanghai & 148             & 2800                &  5.29 \%                 & 03/01/14
\\[0.1 cm] \hline
\end{tabular}
\end{table}

As previously discussed, while we consider Manhattan part of the trip-level datasets, taxi trips are recorded along with taxi IDs. This means we can find $N_V^*$ for this data set (as opposed to only $N_T^*$). The average number of trips per taxi is X, when remarkably translates to $N_V^* = 30$: just 30 random taxis cover half of the scannable street segments. Even more remarkably, one-third of the scannable street segments are scanned by just five random taxis! Supplementary Figure~\ref{manhattan_segments} displays these figures.


\section*{Supplementary Note 5}

\textbf{Minimum street sampling problem}. In the main text we quantified the sensing potential of a vehicle fleet by $\langle C \rangle_{(N_V, m)}$, the average number of segments covered $m$ times when $N_V$ randomly selected vehicles were equipped with a sensor. Note that here the number of vehicles $N_V$ was the independent variable. In some contexts, it might be advantageous to have the reverse scenario, in which $C$ is the independent variable. That is, given a target coverage $\bar{C}$, how many vehicles are needed to ensure this coverage is attained (with a target probability $\bar{p}$). We can this `minimum street sampling' problem, and define and solve it below.

\begin{definition}
(MINIMUM STREET SAMPLING): Given a street network $S$, an observation period $T$, a minimum sampling requirement $m$, and a collection $\mathcal{V}$ of vehicles moving in $S$ during $T$, where vehicle trajectories are taken from $\mathcal{P}$ according to a given probability distribution \textbf{P}; what is the minimum number $N_V^*$ of vehicles randomly selected from $\mathcal{V}$ such that $\mathbb{P}(C(N_V,m) \geq \bar{C}) \geq \bar{p}$, where $0 < \bar{C} \leq 1$ is the target street coverage and $\bar{p}$ is a target probabilistic sampling guarantee?
\end{definition}

This formulation of the sensing potential problem is likely of more utility for urban managers, wishing to know how many vehicles to equip to sensors to guarantee a certain coverage. The minimum street sampling problem is harder to solve that the 'sensing potential of a fleet' problem. This is because it requires the survival function of the multinomial distribution $\mathbb{P}_{N_T}(M_1 \geq m_1, M_2 \geq m_2 \dots,)$, which to our knowledge has no known closed form. We here adapt a technique used in \cite{levin1981representation} to derive an excellent approximation to this survival function.


\textbf{Approximation of survival function}. The pdf for the multinomial distribution is 
\begin{equation}
\mathbb{P}_{N_T}(M_1 = m_1, M_2 = m_2, \dots) = \frac{N_T!}{m_1! \dots m_{N_S}!} \prod_k^{N_S} p_k^{m_k}
\end{equation}
where $N_T$ is the number of balls which have been dropped, $N_S$ is the number of bins, $M_i$ is the random number of balls in bin $i$, and $p_i$ is the probability of selecting bin $i$. We seek the survival function
\begin{equation}
\mathbb{P}_{N_T}(M_1 \geq m_1, M_2 \geq m_2, \dots).
\end{equation}
The idea is to represent each $M_i$ as an independent Poisson random variable, conditional on their sum being fixed (this is a well known identity between the Multinomial and Poisson distributions). First let $A_i$ be the event $X_i \geq m_i$, where $X_i \sim Poi(s p_i)$, where $s$ is a real number (we will explain its significance later). Using Bayes' Theorem, we express the survival function as 
\begin{equation}
\mathbb{P}_{N_T} \Big( A_1,\dots, A_{N_S}| \sum_{i=1}^{N_S} X_i = N_T \Big) = \frac{\mathbb{P}(A_1 \dots, A_{N_S})}{\mathbb{P}(\sum_{i=1}^{N_S} X_i = N_T )} \mathbb{P} \Big(\sum_{i=1}^{N_S} X_i = N | A_1, \dots, A_{N_S} \Big).
\label{bayes}
\end{equation}
The numerator in the first term is easily found, since the events $A_i$ are independent Poisson random variables. Recalling that if $X_i \sim Poi(\lambda_i)$ then $\mathbb{P}(X_i \geq m_i) = 1-\Gamma (m_i,\lambda_i )/\Gamma (m_i) $, where $\Gamma(n,x) = \int_x^{\infty} t^{n-1} e^{-t} dt$ is the upper incomplete gamma function, we find 
\begin{equation}
\mathbb{P}(A_1 \dots, A_{N_S}) = \prod_{i=1}^{N_S} \Bigg( 1-\frac{\Gamma (m_i, s p_i )}{\Gamma (m_i)} \Bigg).
\end{equation}
The denominator is also easy to find. Since $X_i \sim Poi(s p_i)$ and  $\sum_i p_i = 1$, we see $\sum_i X_i \sim Poi(s)$ (sums of poissons are also poisson). Then 
\begin{equation}
\mathbb{P} \Big( \sum_{i=1}^{N_S} X_i = N_T \Big) = \frac{s^N_T e^{-s}}{N_T!}
\end{equation}
For the second term in \eqref{bayes}, we note that conditioning on the joint event $A_1, A_2, \dots$ means the range of the summands are constrained to $[a_i, \infty]$. Hence the summands, which we call $Y_i$, are truncated Poisson random variables, which we denote by $ Y_i \sim Poi_{[a_i, \infty]}(s p_i)$. We note that the mean of a truncated Poisson random variable is not the same as an untruncated one. In particular, if $W_i \sim Poi_{[a,\infty]}(\lambda)$, then
\begin{align}
\mathbb{E} (W_i) &= \lambda \frac{q_{a-1}}{q_{a}} \\
Var(W_i) &= \lambda^2 \frac{ q_{a-2} q_{a} - q_{a-1}^2 }{q_{a}^2} + \lambda \frac{q_{a-1}}{q_{a}}
\end{align}
where
\begin{align}
q_{a} = \left\{
        \begin{array}{ll}
            1 - \frac{\Gamma(a,\lambda)}{\Gamma(a)} & \quad a \geq 1 \\
            0 & \quad a < 1
        \end{array}
    \right.
\end{align}

\noindent
Returning to the second term in \eqref{bayes}, we find
\begin{equation}
\mathbb{P} \Big( \sum_{i=1}^{N_S} X_i = N_T | A_1, \dots, A_{N_S} \Big) = \mathbb{P} \Big( \sum_{i=1}^{N_S} X_i = N_T | A_1, \dots, A_{N_S} \Big) =  \mathbb{P} \Big( \sum_{i=1}^{N_S} Poi_{[a_i, \infty]}(s p_i) = N_T  \Big).
\end{equation}
We were unable to find an analytic form for the above sum. Instead, we used a first order normal approximation. This states that for a sequence of random variables $(W_i)_i$ with mean $\mu_i$ and variance $\sigma_i^2$,
\begin{equation}
\sum_{i=1}^{N_S} W_i \xrightarrow[]{d} N(s_{\mu},s_{\sigma}) 
\end{equation}
as $N_s \rightarrow \infty$, where 
\begin{align}
s_{\mu} = \sum_i \mu_i \\
s_{\sigma}^2 = \sum_i \sigma_i^2.
\end{align}
Then the term becomes
\begin{equation}
\mathbb{P} \Big( \sum_{i=1}^{N_S} X_i = N_T | A_1, \dots, A_{N_S} \Big) = \frac{1}{\sqrt{2 \pi} s_{\sigma}} e^{ - \frac{(N_T - s_{\mu})^2}{2 s_{\sigma}^2} }
\end{equation}
Pulling all this together
\begin{equation}
\mathbb{P}_{N_T}(M_1 \geq m_1, M_2 \geq m_2, \dots) \approx \frac{N_T!}{s^{N_T} e^{-s}} \frac{1}{\sqrt{2 \pi} s_{\sigma}} e^{ - \frac{(N_T - s_{\mu})^2}{2 s_{\sigma}^2} }  \prod_{i=1}^{N_S} \Bigg( 1-\frac{\Gamma (m_i, s p_i )}{\Gamma (m_i)} \Bigg)\label{temp2}
\end{equation}

\noindent
Now, the variable $s$ is a free parameter. Determining the optimal $s$ is an open problem. Following \cite{levin1981representation} we use $s = N_T$, which, when inserted into \eqref{temp2}, along with Stirling's approximation $\frac{N_T!}{N_T^{N_T} e^{-N_T}} \approx \sqrt{2 \pi N_T}$ yields our final expression

\begin{equation}
\mathbb{P}_{N_T}(M_1 \geq m_1, M_2 \geq m_2, \dots) \approx \sqrt{\frac{N_T}{s_{\sigma}^2}} e^{ - \frac{(N_T - s_{\mu})^2}{2 s_{\sigma}^2} }  \prod_{i=1}^{N_S} \Bigg( 1-\frac{\Gamma (m_i, N_T p_i )}{\Gamma (m_i)} \Bigg). \label{final}
\end{equation}
\noindent
To test the accuracy of the above approximation to the survival function, we compared it to Monte Carlo estimates. The results are shown in Figure~\ref{approx}, in which excellent agreement is evident.


\textbf{Solve minimum street sampling}. We leverage the survival function \eqref{final} to solve the minimum street sampling problem in the same way as we did to solve for $C$ in the main text: we assume placing $N_T$ trajectories of random length $L$ into $N_S$ bins is the same as placing $L*N_T$ balls into $N_S$ bins, 
\begin{align}
\mathbb{P}_{(N_T, L)}(M_1 \geq m_1, \dots) = \sum_{n=0}^{\infty} & \mathbb{P}_{(N_T, L=1)}(M_1 \geq m_1, \dots) \mathbb{P}(S_{N_T} = n)
\label{final_P1}
\end{align}
\noindent
where $\mathbb{P}_{(N_T,L=1)}(M_1 \geq m_1, \dots)$ is given by equation \eqref{final}. As for the expression for $C$, this can be extended to the vehicle level by replacing $L$ by $B$. Also as in the main text, this sum is well dominated by its average, leading to the simpler expression
\begin{align}
\mathbb{P}_{(N_T, L)}(M_1 \geq m_1, \dots) = \mathbb{P}_{(\langle L \rangle *N_T, L=1)}(M_1 \geq m_1, \dots)
\label{final_P}
\end{align}
\noindent
When full coverage $\bar{C} = 1$ is desired, equation \eqref{final_P} solves the minimum street sampling problem. However, when less than full coverage $C < 1$ is desired, we must marginalize over all combinations of $N_S*C$ segments above threshold. This is because in our formulation of the minimum street sampling problem we require just a bare fraction $\bar{C}$ of segments be covered, which is achievable by a large number of combinations of segments. Of course if \textit{targeted} coverage were desired (i.e were specific street segments were desired to be senses with specific sensing requirements $m$), then \eqref{final_P} could be used. Staying within our current formulation however, an enumeration of all $C N_S$ combinations of bins is required to marginalize $\mathbb{P}(M_1 \geq m_1, \dots)$. For large $N_S$ enumerating these combinations is infeasible. To get around this, we instead estimate $\mathbb{P}_{(\langle L \rangle *N_T, L=1)}(M_1 \geq m_1, \dots)$ by Monte Carlo; we draw samples of size $\langle L \rangle *N_T$ from a multinomial distribution 1000 times, and count the fraction of times at least $\bar{C}$ of the $N_S$ bins are above the threshold $m$. This lets us estimate $\mathbb{P}(C > \bar{C})(N_T)$, from which we can read off the desired $N_T^*(\bar{P})$ solving the minimum street sampling problem.

In Figure~\ref{P_all_cities} we compare our predictions versus data for a target coverage of $\bar{C} = 0.5$. While the precise shapes of the theoretical and empirical curves do not agree, our model correctly captures the right range of variation. In particular, the error $N_{T,model}(\bar{P} \approx 1)$ - $N_{T,data}(\bar{P} \approx 1)$ is $\approx 200$. Expressed relative to the total number of trips, this is $ \sim 10^{-4}$ for the NYC and Singapore datasets, and $\sim 10^-2$ for the other datasets, which is good accuracy.
